\documentclass[aps,prd,showpacs,showkeys,twocolumn,nofootinbib]{revtex4-1}
\usepackage{epsfig}
\usepackage[colorlinks,linkcolor=blue,anchorcolor=blue,citecolor=blue,urlcolor=blue,breaklinks=true]{hyperref}
\usepackage{graphics}
\usepackage{color}

\begin{document}
\author{Bo-Lin Li$^{1,3}$}~\email[]{Email: blli@nju.edu.cn}
\author{Zhu-Fang Cui$^{2,3,4,}$}~\email[]{Email: phycui@nju.edu.cn}
\author{Bo-Wen Zhou$^{1,3}$}
\author{Sun An$^{1,3}$}~\email[]{Email: sunan@nju.edu.cn}
\author{Li-Ping Zhang$^{3,5}$}
\author{Hong-Shi Zong$^{2,3,4,6}$}

\address{$^{1}$College of Engineering and Applied Sciences, Nanjing University, Nanjing, Jiangsu 210093, China}
\address{$^{2}$Department of Physics, Nanjing University, Nanjing, Jiangsu 210093, China}
\address{$^{3}$Nanjing Proton Source Research and Design Center, Nanjing, Jiangsu 210093, China}
\address{$^{4}$State Key Laboratory of Theoretical Physics, Institute of Theoretical Physics, CAS, Beijing, 100190, China}
\address{$^{5}$Key laboratory of road construction \& equipment of MOE, Chang'an University, Xi'an, Shanxi 710064, China}
\address{$^{6}$Joint Center for Particle, Nuclear Physics and Cosmology, Nanjing, Jiangsu 210093, China}

\title{Finite volume effects on the chiral phase transition from Dyson-Schwinger equations of QCD}
\begin{abstract}

Within the framework of Dyson-Schwinger equations of QCD, we study the finite volume effects on the chiral phase transition, especially the influence on the position of the possible pseudo-critical end point (pCEP). The results show that in the chiral limit case (the current quark mass $m=0$), the absolute value of quark condensate decreases for smaller volumes, and more interestingly, so does the pseudo-critical temperature $T_c(\mu=0)$, which is in agreement with the Polyakov Nambu--Jona-Lasinio model result and opposite to the Polyakov linear sigma model prediction. These conclusions hold for $m>0$ case in our calculations. Moreover, the results of pCEP as a function of different volumes show that $T$ of pCEP also decreases for smaller volumes, but $\mu$ of pCEP will increase, which are qualitatively more close to Polyakov linear sigma model results. For our model setup, results for systems with a size larger than (5 fm)$^3$ closely approximate those from infinite volume, but if the volume is smaller, the corrections are non-negligible, even significantly affect signatures of the results from an infinite system. There also exists some possibility that, if the system size is too small, the whole phase transition would be crossover, which means no pCEP exists at all. It is no doubt that, finite volume effects deserve further researches.

\end{abstract}



\maketitle

\section{INTRODUCTION}
Quantum ChromoDynamics (QCD), which is the underlying theory of strongly interacting quarks and gluons, is part of the Standard Model of particle physics. However, QCD also shows quite special complexity and nonlinear properties in the low energy field, consequently, two of its most fundamental characters: dynamical chiral symmetry breaking (DCSB) and color confinement, have not been fully understood up to now. DCSB is responsible for about 98\% masses of the luminous universe, while color confinement is one of the fundamental puzzles of the nature. Understanding the generation of masses, the reason of color confinement, the phases of QCD at extreme temperatures (about 0.15 GeV $\sim$ 1.7$\times$10$^{12}$ K, which is almost 10 orders of magnitude hotter than the surface of the Earth and still more than 5 orders of magnitude hotter than the center of the Sun! This is supposed to bring us back to the evolution of the early universe within 10 $\mu s$ after the Big Bang) and/or densities (about 1.6$\times$10$^{33}$ Pa $\sim$ 0.01 GeV/fm$^3$) as well as their transitions are some of the hot topics and great challenges in high energy physics. As for the nature of the QCD phase transitions, a popular scenario favors a crossover (or second order for the chiral limit case, where the current quark mass $m$=0) at small chemical potential $\mu$, and then a first-order chiral transition for larger chemical potential at a critical end point (CEP, or tri-critical point for the chiral limit case). The search for the position or even the existence of such a CEP is also one of the main goals for the theoretical and experimental physics~\cite{Cui:2013tva,*Cui:2013aba,*Cui:2018bor,Ayala:2014jla,Fischer:2014ata,Xin:2014dia,Eichmann:2015kfa,Braguta:2015owi,
Inagaki:2015lma,Kovacs:2016juc,Gao:2016qkh,Shao:2016fsh,Wang:2016fzr,Cui:2017ilj}. However, it is not yet clarified directly from the first principles of QCD.

On the other hand, the matters we want to describe, which are formed by interacting particles, obviously have a finite volume (size). Finite volume effects also play or may play important roles in systems ranging from the physics of ultracold atom clouds or optical lattices over condensed matter systems, multi-layer systems to heavy-ion collisions where the interaction region is relatively sharply bounded, see Refs.~\cite{Ma:1993qz,Ladrem:2004dw,hep-ph/0512233,0912.3686,1210.8033,Bhattacharyya:2014uxa,Fister:2015eca,Pan:2016ecs}, also a recent review Ref.~\cite{Klein:2017shl}. For example, the radii of possible quark gluon plasma (QGP) are estimated to be 2 to 10 fm, depending on the size of the colliding nuclei, the center of mass energy, and the centrality of collisions, etc. Lattice QCD (lQCD), which is thought to be the most reliable non-perturbative QCD method at present, and other theoretical approaches with spacetime discrete are also related to finite volume effects due to finite number of lattice or grid points. Accordingly, to get a fundamental description of the thermodynamic phases that be created in the experiments, it is imperative to have a clear understanding of the finite volume effects. In general, from statistical physics we know that phase transitions are infinitely sharp only in the thermodynamic limit (namely, the size of the system $V\rightarrow\infty$), and only in this limit is the thermodynamical potential or any of its derivatives singular at some critical point. Also, finite size scaling analysis tells us that we can extract the true critical behaviors of infinite systems from calculations about finite systems, by studying how the characteristic quantities vary with the size of the system. Besides some apparent diversity in the underlying structure of systems undergoing phase transitions, there are usually some universal global behaviors, which only depend on the range of interaction and the dimensionality of the related system, which then ensures the singular behaviour of some characteristic observables near criticality identical for different systems if appropriately scaled. The universal behavior is usually characterized by indices named critical exponents.

However, we notice that at present effective model studies of the finite volume effects do not always agree with each other, among which, let us focus on the behaviors of the QCD chiral phase transition, especially the position of pCEP~\footnote{As we discussed above, only in the thermodynamic limit will phase transitions be infinitely sharp, and then there exist some critical point with divergence. The ``CEP'' for a finite volume is just the point connect crossover and first order phase transition line, which has a very large but not infinity susceptibility, then is not second order. This is why we call it ``pseudo-CEP''.} , where Polyakov linear sigma model (PLSM) and quark-meson model find that pCEP is shifted towards higher values of the chemical potential and lower values of the temperature quickly for smaller volumes~\cite{Palhares:2009tf,Tripolt:2013zfa,Magdy:2015eda}, while Polyakov Nambu--Jona-Lasinio model (PNJL) gives a similar result for the temperature of pCEP but the corresponding chemical potential is almost stable~\cite{Bhattacharyya:2012rp}. More interestingly, for the pseudo-critical temperature $T_c(\mu=0)$, PLSM predicts that it increases for smaller volumes, but PNJL model gives the opposite result! Based on our experience on studies related to chiral chemical potential $\mu_5$~\cite{Wang:2015tia,Xu:2015vna,Cui:2016zqp}, namely, the results from Dyson-Schwinger equations (DSEs)~\cite{PPNP.33.477,alkofer2001infrared,Fischer:2006ub,PPNP.77.1-69} usually have better (at least qualitatively) agreement with lQCD calculations while we should be very careful when using chiral models, we will discuss the finite volume effects on the QCD chiral phase transition within DSEs in this work, especially the evolution of pCEP.  The following of this paper is organized as this: in Sect.~\ref{dsefv} we give a brief introduction to the DSEs at finite temperature and nonzero chemical potential, and with the help of a scalar susceptibility we also discuss the nature of the chiral phase transition; then in Sect.~\ref{cepfv}, we discuss the influences of the finite volume effects on the chiral phase transition, and mostly focus on the behaviour of the pCEP; at last, a brief summary is given in Sect.~\ref{sum}.

\section{Dyson-Schwinger equations and the QCD chiral phase transition}\label{dsefv}
QCD is asymptotically free, which means the coupling decreases with increasing energy. Two main characteristics of QCD: color confinement and DCSB, are non-perturbative phenomena whose precise understanding continues to be a hot topic of theoretical and experimental researches. Also, the theory toolkit to study QCD matter is quite diverse, thanks to its complexity in low energy region and the rich phenomena it describes. Here we review the basic formula of Dyson-Schwinger equations, which is widely used in the non-perturbative region of QCD. At zero temperature and zero chemical potential, the DSE of the quark propagator reads~\cite{PPNP.33.477} (we will always work in Euclidean space, and take the number of flavors $N_f=2$ while the number of colors $N_c=3$ throughout this paper. Moreover, as we employ a ultra-violet finite model, renormalization is actually unnecessary)
\begin{equation}
S(p)^{-1}=S_0(p)^{-1}+\frac{4}{3} \int\frac{\mathrm{d}^4q}{(2\pi)^4}g^2D_{\mu\nu}(p-q)\gamma_\mu S(q)\Gamma_\nu,\label{dse1}
\end{equation}
where $S(p)^{-1}$ is the inverse of the dressed quark propagator ($p$ and $q$ are momenta),
\begin{equation}
S_0(p)^{-1}=i\gamma\cdot p+m,
\end{equation}
is the inverse of the free quark propagator, $g$ is the coupling constant of strong interaction, $D_{\mu\nu}(p-q)$ is the dressed gluon propagator, and $\Gamma_\nu=\Gamma_\nu(p,q)$ is the dressed quark-gluon vertex. According to the Lorentz structure analysis, we have
\begin{equation}
S(p)^{-1}=i{\not\!p}A(p^2)+B(p^2),\label{ppg1}
\end{equation}
where $A(p^2)$ and $B(p^2)$ are scalar functions of $p^2$. Once the gluon propagator and the quark-gluon vertex are specified, this equation can then be solved numerically.

The extension of the above quark DSE to the nonzero temperature and nonzero quark chemical potential version is systematically accomplished by transcription of the quark four-momentum via $p\rightarrow p_k=(\vec{p}, \tilde\omega_k)$, where $\tilde\omega_k=\omega_k+i\mu$ with $\omega_k=(2k+1)\pi T$, $k\in Z\!\!\!\!Z$ the fermion Matsubara frequencies, and no new parameters are introduced
\begin{eqnarray}\label{dse2}
S(p_k)^{-1}=S_0(p_k)^{-1}+\frac{4}{3}T \int\!\!\!\!\!\!\!\!\sum g^2D_{\mu\nu}(p_k-q_n)\gamma_{\mu}S(q_n)\Gamma_{\nu}.\nonumber\\
\end{eqnarray}
where
\begin{equation}\label{free}
S_0(p_k)^{-1}=i\gamma \cdot p_k+m,
\end{equation}
and $\int\!\!\!\!\!\!\!\sum$ denotes $\sum_{l=-\infty}^{+\infty}\int\frac{\mathrm{d}^3\vec{q}}{(2\pi)^3}$. Nevertheless, its solution now should have four independent amplitudes due to the breaking of $O(4)$ symmetry down to $O(3)$ symmetry
\begin{eqnarray}
S(p_k)^{-1}=&&i\not\!\vec{p}\,A(\vec{p}\,^2,\tilde\omega_k^2) + \mathbf{1}B(\vec{p}\,^2,\tilde\omega_k^2)\nonumber\\
 &&+i\gamma_4\,\tilde\omega_kC(\vec{p}\,^2,\tilde\omega_k^2)+\not\!\vec{p}\,\gamma_4\,\tilde\omega_kD(\vec{p}\,^2,\tilde\omega_k^2),
\end{eqnarray}
where $\not\!\!\vec{p}=\vec{\gamma}\cdot\vec{p}$, $\vec{\gamma}=(\gamma_1, \gamma_2, \gamma_3)$, and the four scalar functions ${\cal F}= A$, $B$, $C$, $D$ are complex and satisfy
\begin{equation}
{\cal F}(\vec{p}\,^2,\tilde\omega_k^2)^\ast ={\cal F}(\vec{p}\,^2,\tilde\omega_{-k-1}^2)\,.\label{abc}
\end{equation}
Here the dressing function $D$ is power-law suppressed in the ultra-violate region, so that actually does not contribute in all cases investigated in our work. At zero temperature but nonzero chemical potential case, $D$ vanishes exactly since the corresponding tensor structure has the wrong transformation properties under time reversal~\cite{ZPA.352.345--350}, so in most cases we can just neglect $D$, and get the commonly used general structure of the inverse of quark propagator as
\begin{equation}
S(p_k)^{-1}=i\not\!\vec{p}\,A(\vec{p}\,^2,\tilde\omega_k^2) + \mathbf{1}B(\vec{p}\,^2,\tilde\omega_k^2)+i\gamma_4\,\tilde\omega_kC(\vec{p}\,^2,\tilde\omega_k^2).\label{ppg3}
\end{equation}

For the dressed-gluon propagator, the general form is,
\begin{equation}\label{mtgluon}
g^2D_{\mu\nu}(k_{nl})=P_{\mu\nu}^TD_T(\vec{k}^2,\omega_{nl}^2)+P_{\mu\nu}^LD_L(\vec{k}^2,\omega_{nl}^2),
\end{equation}
where $k_{nl}=(\vec{k},\omega_{nl})=(\vec{p}-\vec{q},\omega_n-\omega_l)$, $P_{\mu\nu}^{T,L}$ are transverse and longitudinal projection operators, respectively. For the domain $T<0.2$ GeV that we are concerned in this work, it has been proved that $D_T=D_L$ is a good approximation~\cite{Cucchieri:2007ta}. For the in-vacuum interaction, in this work we will adopt the following form of $Ansatz$~\cite{PRL.106.172301},
\begin{equation}\label{simMT}
D_T=D_L=D_0\frac{4\pi^2}{\sigma^6}k_{nl}^2e^{-k_{nl}^2/\sigma^2},
\end{equation}
which is a simplified version of the famous as well as widely used one in Refs.~\cite{PhysRevC.56.3369,PhysRevC.60.055214}. It can be proved that this dressed gluon propagator at $T=0$ violates the axiom of reflection positivity~\cite{glimm1987quantum}, and is therefore not observable, in other words, the excitation it describes is confined. The same is true for the dressed quark propagator too. In fact, we can take the gluon propagator as input, and then solve the quark propagator numerically, the results show that there is no singularity on the real timelike, $p^2$ axis~\cite{Windisch:2016iud}, which implies that quarks are indeed confined. For the quark-gluon vertex, we will take the rainbow truncation in this work, which means a simple but symmetry-preserving bare vertex is adopted,
\begin{equation}\label{rainbow}
\Gamma_\nu(p_n,q_l)=\gamma_\nu.
\end{equation}

\begin{figure}
\includegraphics[width=0.45\textwidth]{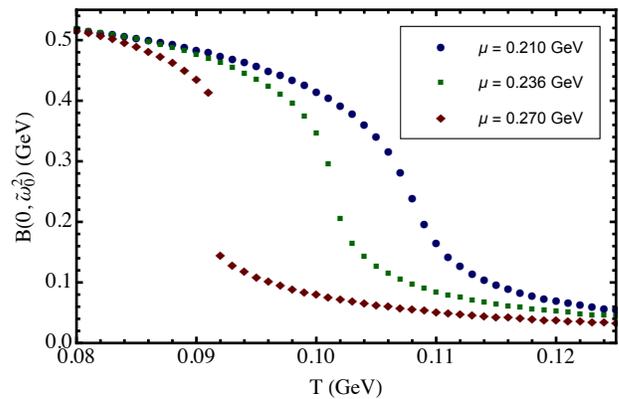}
\caption{(color online) $B(0,\tilde\omega_0^2)$ as a function of $T$ for three different $\mu$ (for $u$ and $d$ quarks).}\label{bmu}
\end{figure}

\begin{figure}
\includegraphics[width=0.45\textwidth]{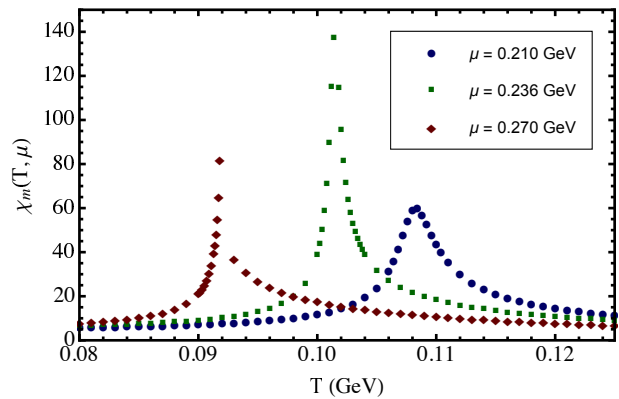}
\caption{(color online) $\chi_m(T,\mu)$ as a function of $T$ for three different $\mu$ (for $u$ and $d$ quarks).}\label{sus}
\end{figure}

The related parameters, $D_0$ and $\sigma$ are usually fixed by fitting some observables: the pion decay constant ($f_\pi=0.095$ GeV) and the pion mass ($m_\pi=0.139$ GeV). In this work we use $D_0$=1.0 GeV$^2$ and $\sigma=0.5$ GeV. For the current quark mass we will use $m$=0.005 GeV. Now substituting Eqs.~(\ref{free}), (\ref{ppg3}), (\ref{simMT}), and (\ref{rainbow}) into Eq.~(\ref{dse2}), we can solve the quark DSE for different temperature and chemical potential by means of numerical iteration. As an example, we show $B(0,\tilde\omega_0^2)$ as a function of $T$ for different $\mu$ in Fig.~\ref{bmu}. On the other hand, the vacuum of QCD is also difficult to characterize, one possible way is using several non-perturbative objects, such as various susceptibilities, which are the linear responses of the QCD condensate (such as the quark condensate $\langle\bar\psi\psi\rangle$, which is also called as chiral condensate, since it is the order parameter for the chiral phase transition in the chiral limit) to various external fields~\cite{Cui:2015xta}. Here the chiral susceptibility with respect to $m$, which is defined as
\begin{equation}\label{chim}
\chi_m(T,\mu)=\frac{\partial B(0,\tilde\omega_0^2)}{\partial m},
\end{equation}
is plotted in Fig.~\ref{sus}.

It can be seen from Fig.~\ref{bmu} that the scalar function $B(0,\tilde\omega_0^2)$ will decrease when the temperature $T$ increase, this phenomenon holds to be true for the chemical potential $\mu$ and momentum $\vec{p}^2$ too. It is known that the scalar part $B(\vec{p}\,^2,\tilde\omega_k^2)$ of the quark propagator Eq.~(\ref{dse2}) in some sense reflects the dressing effect of the quark, so the results show that the dressing effect becomes weaker and weaker for higher $T, \mu$ and $\vec{p}^2$. We can also see from Fig.~\ref{bmu} that for different values of $\mu$, $B(0,\tilde\omega_0^2)$ may behave qualitatively different: for $\mu$ lower than a critical value $\mu_{\rm c}=0.236$ GeV for this work, $B(0,\tilde\omega_0^2)$ change gradually but continuously from the Nambu solution\footnote{Corresponds to the Nambu (or Nambu-Goldstone) phase, where DCSB and color confinement of quarks and gluons are the key emergent phenomena.} to the Wigner solution\footnote{The opposite concept to Nambu solution, which corresponds to the Wigner (or Wigner-Weyl) phase, where the chiral symmetry is (partially) restored, and is then related to the theoretically predicted QGP.}; while for $\mu$ higher than $\mu_{\rm c}$, there will appear a sudden discontinuity at some critical $T$, which indicates there will be a first order phase transition nearby. To determine the nature of the chiral phase transition, especially the critical values of $\mu$ and $T$, are often determined via various susceptibilities. As we can see from Fig.~\ref{sus}, for $\mu\leq \mu_c$, the susceptibility $\chi_m$ indicate a crossover from the Nambu phase to the Wigner phase, and the peak will grow higher and higher when $\mu$ increase to $\mu_c$. At $\mu_c$, $\chi_m$ shows a quite sharp and narrow divergent peak, which demonstrates a second-order phase transition here that corresponds to the CEP. And for $\mu\geq \mu_c$, first order phase transition will take place. Based on these results, we can then study the finite volume effects to the chiral phase transition, especially the corresponding behavior of pCEP.

\section{Finite volume effects on the chiral phase transition}\label{cepfv}
As discussed above, studies of the QCD phase structure have been a hot topic for decades, while most studies assume an infinite volume of the system. However, it is not obvious that the size of what we concern is large enough to apply the thermodynamic limit. Especially, if the size of the system is small, we then must take the deviations from thermodynamic calculations into account. For instance, the fluctuations of order parameters, induced by the finite volume effects, may lower the critical temperature even change the order of the phase transition. Large fluctuations near phase transitions may also invalidate the mean-field approximation.

To include the effects of finite volume, in principle we should choose proper boundary conditions (anti-periodic for fermions while periodic for bosons), which would lead to an infinite sum over discrete momentum values $p_i=\pi n_i/R$, where $i=x,y,z$, $n_i$ are all positive integers, and $R$ is the lateral size of a cubic volume. Proper effects of surface and curvature also should be incorporated. In this work we will follow the ideas of Ref.~\cite{Bhattacharyya:2012rp}, in which a lower momentum cutoff $p_{min}=\pi/R$ is used as an approximation, and a few other simplifications:

\noindent
(i) The infinite sum will be replace by an integration over a continuous variation of momentum albeit with the lower cutoff. Notice that DSEs does not need any hard momentum cutoff like PNJL model, the calculations will no doubt be more reliable for small volumes than PNJL model;

\noindent
(ii) Neglect the surface and curvature effects;

\noindent
(iii) We neglect the modifications to the vacuum mean-field parameters due to finite volume effects. The philosophy is to hold the known physics at zero $T$, zero $\mu$ and infinite volume, and treat $R$ as a thermodynamic variable in the same footing as $T$ and $\mu$.

Therefore, any variation due to change in either of these thermodynamic parameters were translated into the changes in the effective masses of quarks, and through them to other quantities.

\begin{figure}
\centering
\includegraphics[width=0.47\textwidth]{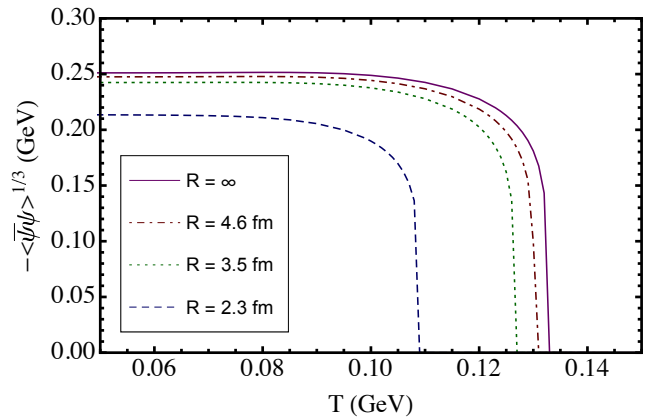}
\caption{(color online) Volume dependent chiral condensate as a function of $T$ for the chiral limit case, where $\mu$ is fixed to be 0 (for $u$ and $d$ quarks).}\label{con}
\end{figure}

Here in Fig.~\ref{con}, we show the different volume results of chiral condensate as a function of $T$ for the chiral limit case, where $\mu$ is fixed to be 0, and obvious phase transitions are observed. We see that $\langle\bar\psi\psi\rangle$ does have volume dependence: its absolute value decreases for smaller volumes. More interestingly, we find that the pseudo-critical temperature $T_c(\mu=0)$ also decrease for smaller volumes, which is in agreement with PNJL model result and opposite to the PLSM prediction. These conclusions also hold for $m>0$ case in our calculation.

\begin{figure}
\centering
\includegraphics[width=0.47\textwidth]{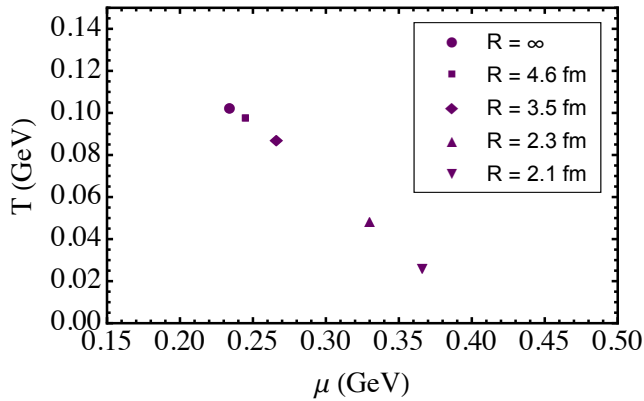}
\caption{Trajectory of volume dependent pCEP (for $u$ and $d$ quarks).}\label{pCEP}
\end{figure}

Furthermore, the results of pCEP as a function of different volumes are given in Fig.~\ref{pCEP}, where $T$ of pCEP also decreases for smaller volumes, but $\mu$ of pCEP will increase, which are qualitatively more close to PLSM results. We also notice that, for systems with a size larger than (5 fm)$^3$, infinite volume approximation would be regarded as acceptable within our model setup, but if the volume is smaller, the corrections are clearly non-negligible, and might significantly affect signatures of the pCEP based on estimates from an infinite system. This is an meaningful fact for the CEP search in heavy-ion collision experiments, since if the size of QGP is small, according to our calculation, we should try to find pCEP at low $T$ and high $\mu$ region, which means to collide the ions at lower energy. There also exists some possibility that, if the system size is too small, the whole phase transition will be crossover, which means no pCEP exists at all.  All in all, finite volume effects are meaningful and necessary for further research.

\section{summary}\label{sum}
The QCD chiral phase transition, especially the existence and position of the possible critical end point (CEP), is one of the hot topics and great challenges in high energy physics. The related effects of finite volume are certainly important since realistic systems are of course finite, while phase transitions are known to be infinitely sharp, signaled by some singularities, only in the thermodynamic limit. To this end, the finite volume effects on the chiral phase transition are studied within the framework of Dyson-Schwinger equations of QCD in this work, where we mainly discuss the influence on the position and change of the possible pseudo-CEP (pCEP). For the chiral limit case (the current quark mass $m$ is taken to be zero, so that we can discuss the exact chiral phase transition), we find that the absolute value of quark condensate decreases for smaller volumes, and the pseudo-critical temperature $T_c(\mu=0)$ also decreases for smaller volumes, which is in agreement with the Polyakov Nambu--Jona-Lasinio model result, but opposite to the Polyakov linear sigma model prediction. We have also checked that these conclusions are also true beyond the chiral limit. Moreover, the results of pCEP dependence of different volumes show that $T$ of pCEP also decreases for smaller volumes, but $\mu$ of pCEP will increase, which are qualitatively more close to the Polyakov linear sigma model results instead of the Polyakov Nambu--Jona-Lasinio model calculations. From our model calculations, it seems that results for systems with a size larger than (5 fm)$^3$ are closely to those from infinite volume, then the thermodynamic limit will be a good approximation; but if the volume is smaller, the corrections are then non-negligible, even significantly affect signatures of the results from an infinite system. Our results also show that there exists some possibility that, if the system size is smaller than some critical value, the whole QCD chiral phase transition would be crossover, which means no first order phase transition will happen, then no pCEP would exist at all. It is no doubt that, finite volume effects deserve further researches.

\acknowledgments
This work is supported by the National Natural Science Foundation of China (under Grant No. 11805097 and No. 51405027), the Jiangsu Provincial Natural Science Foundation of China (under Grant No. BK20180323), the Fundamental Research Funds for the Central Universities(under Grant No. 021314380035, No. 020414380051, and No. 020414380074), the China Postdoctoral Science Foundation (under Grant No. 2015M581765, and No. 2018M642204), and Open Research Foundation of State Key Lab. of Digital Manufacturing Equipment \& Technology in Huazhong University of Science \& Technology (under Grant No. DMETKF2015015).

\bibliographystyle{apsrev4-1}
\bibliography{NPB}

\end{document}